\documentclass[12pt,preprint]{aastex}

\shorttitle{Central Structure of M15}
\shortauthors{Baumgardt et al.}

\begin{document}


\title{On the central structure of M15}


\author{Holger Baumgardt\altaffilmark{1},        
        Piet Hut\altaffilmark{2},
        Junichiro Makino\altaffilmark{1},
        Steve McMillan\altaffilmark{3},
        Simon Portegies Zwart\altaffilmark{4}}

\altaffiltext{1}{
        Department of Astronomy, University of Tokyo, 7-3-1 Hongo,
        Bunkyo-ku,Tokyo 113-0033, Japan}

\altaffiltext{2}{
        Institute for Advanced Study, Princeton, NJ 08540, USA}

\altaffiltext{3}{
        Department of Physics, Drexel University, Philadelphia, PA
        19104, USA}

\altaffiltext{4}{
        Astronomical Institute ``Anton Pannekoek,'' University of
        Amsterdam, Kruislaan 403, 1098 SH Amsterdam, The Netherlands}


\begin{abstract}
We present a detailed comparison between the latest observational data
on the kinematical structure of the core of M15, obtained with the
Hubble STIS and WFPC2 instruments, and the results of dynamical
simulations carried out using the special-purpose GRAPE-6 computer.
The observations imply the presence of a significant amount of dark
matter in the cluster core.  In our dynamical simulations, neutron
stars and/or massive white dwarfs concentrate to the center through
mass segregation, resulting in a sharp increase in $M/L$ toward the
center.  While consistent with the presence of a central black hole,
the Hubble data can also be explained by this central concentration of
stellar-mass compact objects.  The latter interpretation is more
conservative, since such remnants result naturally from stellar
evolution, although runaway merging leading to the formation of a
black hole may also occur for some range of initial conditions.  We
conclude that no central massive object is required to explain the
observational data, although we cannot conclusively exclude such an
object at the level of $\sim500-1000$ solar masses.  Our findings are
unchanged when we reduce the assumed neutron-star retention fraction
in our simulations from 100\% to 0\%.

\end{abstract}




\newcommand{\msun}{M_{\odot}}
\def\apgt{\ {\raise-.5ex\hbox{$\buildrel>\over\sim$}}\ }
\def\aplt{\ {\raise-.5ex\hbox{$\buildrel<\over\sim$}}\ }

\section{Introduction}

\citet{Gerssenetal2002} have recently reported evidence for an
intermediate-mass ($1.7\pm2.7\times10^3\msun$) black hole (IMBH) at
the center of globular cluster M15.  If confirmed, this would be an
exciting and important discovery, and may necessitate a fundamental
change in our understanding of the dynamical evolution of globular
clusters.  To evaluate the need for such a change, we confront the
observations with the most detailed cluster simulations currently
available.

In the standard view \citep{Spitzer1987,MeylanHeggie1997}, globular
clusters are born with relatively low central densities.  Through
two-body relaxation, some of them may reach core collapse, with very
high central stellar density.  If the cluster contains a significant
population ($\apgt10$\%) of primordial binaries, the kinetic energy
released by binary--binary and binary--single-star interactions
eventually halts the contraction of the core and the cluster reaches a
quasi-steady state \citep{GoodmanHut1989} that may endure for
substantially longer than a Hubble time.  If the cluster contains
few primordial binaries, the contraction of the core is halted instead
at much higher density by the formation of binaries through
three-body interactions.  In this case, there is no steady state, and
the core may exhibit gravothermal oscillations
\citep{Bettwieser1984,Makino1997}.

In this picture, the central density of a globular cluster becomes
high only after several gigayears, since it typically takes several
half-mass relaxation times for core collapse to occur.  This view is
observationally well supported, since most Galactic globular clusters
do have sizeable cores \citep{DjorgovskiMeylan1994,Harris1996}.  It is
unlikely that an IMBH could have formed as a result of M15's core
collapse, as present conditions at the cluster center are unsuitable
for runaway stellar collisions to occur
\citep{Lee1987,PortegiesZwartMcMillan2002}.  Alternative possibilities
are that the cluster was initially very compact and that a runaway
merger leading to an IMBH may have occurred
\citep{PortegiesZwartetal1999}, or that an initial seed black hole
grew slowly over a Hubble time via occasional collisions with other
stars \citep{MillerHamilton2002}, forming an IMBH by the present time.

In this paper we compare the M15 observations with direct $N$-body
simulations of star clusters in which stellar evolution and the
effects of the Galactic tidal field are realistically taken into
account \citep{BaumgardtMakino2002}.  In \S2 we describe our cluster
model and in \S3, we present ``observations'' of our model cluster and
compare them with the actual observations of M15.  We briefly
summarize and conclude in \S4.

\section{Model description}

\citet{BaumgardtMakino2002} have performed simulations of star
clusters with up to 131072 (128k) stars, using the NBODY4 code
\citep{Aarseth1999a} on the GRAPE-6 computer \citep{Makinoetal2002}.
Here we concentrate on a member of their ``Family 2.''  Initial
stellar masses were chosen from a \citet{Kroupa2001} mass function
with lower and upper mass limits of 0.1 and 15 $M_{\odot}$.
Primordial binaries were not included.  The initial distribution of
stars was given by a King model with dimensionless central potential
$W_0 = 7$.  The model cluster was placed on a circular orbit at a
distance of 8.5 kpc from the Galactic center.  The Galactic potential
was treated as a singular isothermal sphere with a constant rotation
velocity of 220 km/s.  Stellar evolution was modeled according to
\citet{Hurleyetal2000}.  The initial half-mass radius of the cluster
(with $N=128$k stars and a mass of $7.2\times 10^4$ M$_\odot$) was 7.1
pc; the initial half-mass crossing time was 4.1 Myr.  Core collapse
occurred at $T=12.6$ Gyr, when the remaining cluster mass was
$\sim2\times 10^4$ M$_\odot$.  The calculation, to the point of
complete dissolution, took about 1000 hours computing time on a
4-board, single-host GRAPE-6 system.  Details of the calculation are
described in \citet{BaumgardtMakino2002}.

Note that our 128k-body model still contains far fewer stars than
M15---we cannot yet perform star-by-star simulations of a relatively
large globular cluster.  Rather, we compare nondimensional quantities,
such as the radial dependence of the velocity dispersion, its slope,
$M/L$ etc.  In the next section we present a comparison of the
luminosity and velocity dispersion profiles near the centers of the
two systems.

In the calculations of \citet{BaumgardtMakino2002}, collisions between
stars were not taken into account, and hence massive black holes could
not form.  We have also performed simulations in which stellar
collisions were properly included
\citep{PortegiesZwartetal2001,PortegiesZwartMcMillan2002} and find
that, for initial conditions appropriate for globular clusters, the
neglect of stellar collisions is justified.

\section{Analysis}

\begin{figure}[htbp!]
\plotone{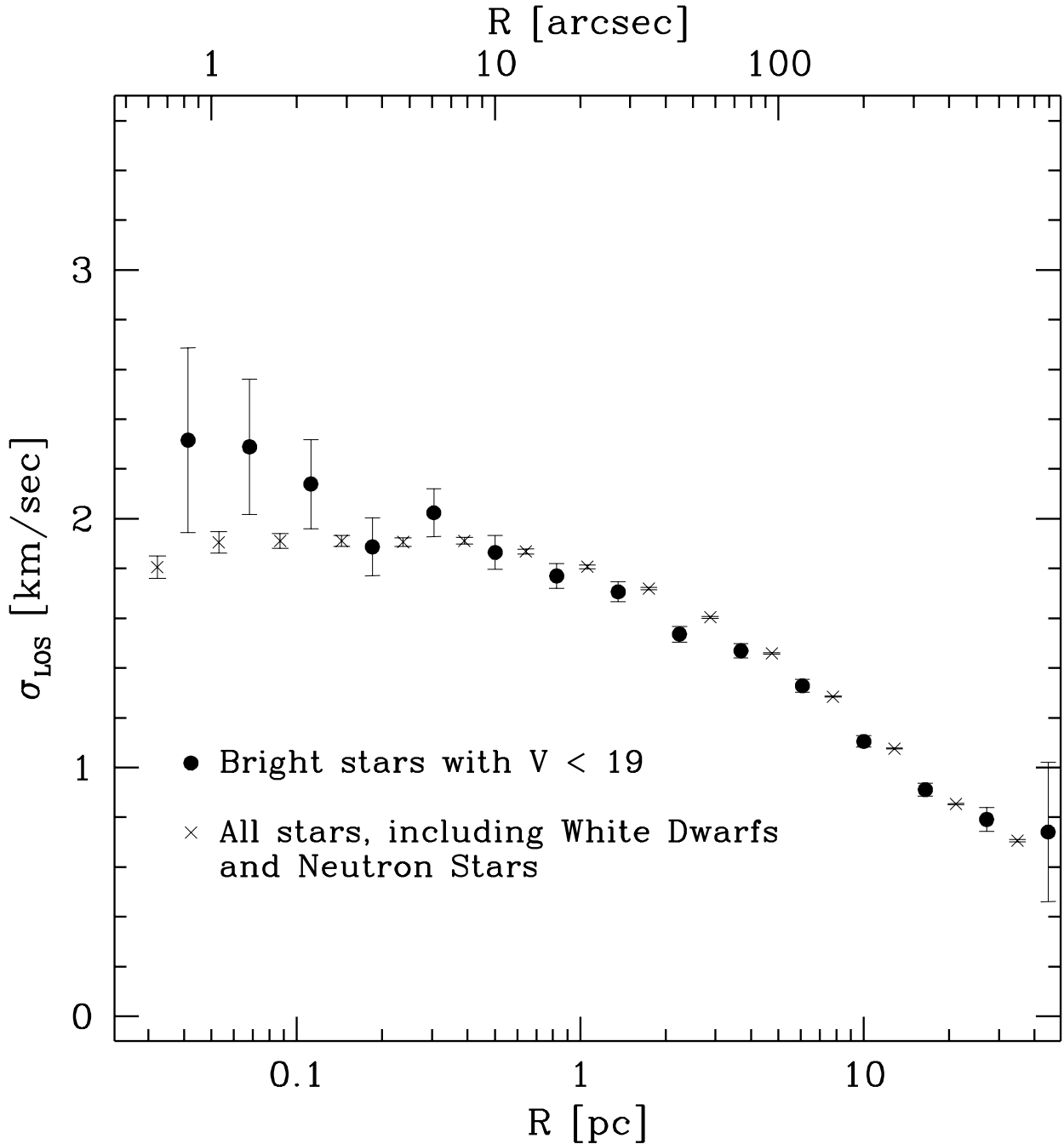}
\caption{Line-of-sight velocity dispersion, $\sigma_{\rm LOS}$, as a
function of projected distance from the cluster center.  Ten snapshots
with time intervals of 50 Myr are overlaid to improve statistics.
Velocity dispersions are averaged over orientation angles.  Crosses
are calculated using all stars in the cluster, and filled circles
using stars with visual magnitude $V < 19$ at 10 kpc.  The upper axis
gives distances in arcseconds, calculated by assuming that our model
cluster is observed from a distance of 10 kpc.
\label{fig1}}
\end{figure}

Figure \ref{fig1} shows the ``observed'' line-of-sight velocity
dispersion profile of our model cluster.  In order to improve
statistics, we have superimposed ten snapshots spanning a 500 Myr
period following core collapse. We calculated the velocity dispersion
of the model cluster in two ways.  First, we determined the velocity
dispersion using all stars (including compact remnants), averaging
over three orthogonal directions.  In the second method, we used only
stars brighter than $V=19$ at the distance of M15 (assumed to be 10
kpc), the sample actually used by \citet{Gerssenetal2002}.  Except for
the innermost parts, both profiles agree rather well with each other;
within the error bars, the model velocity dispersion profile is also
very similar to that of M15 (Figure 9 of Gerssen et al. 2002).

\begin{figure}[htbp!]
\plotone{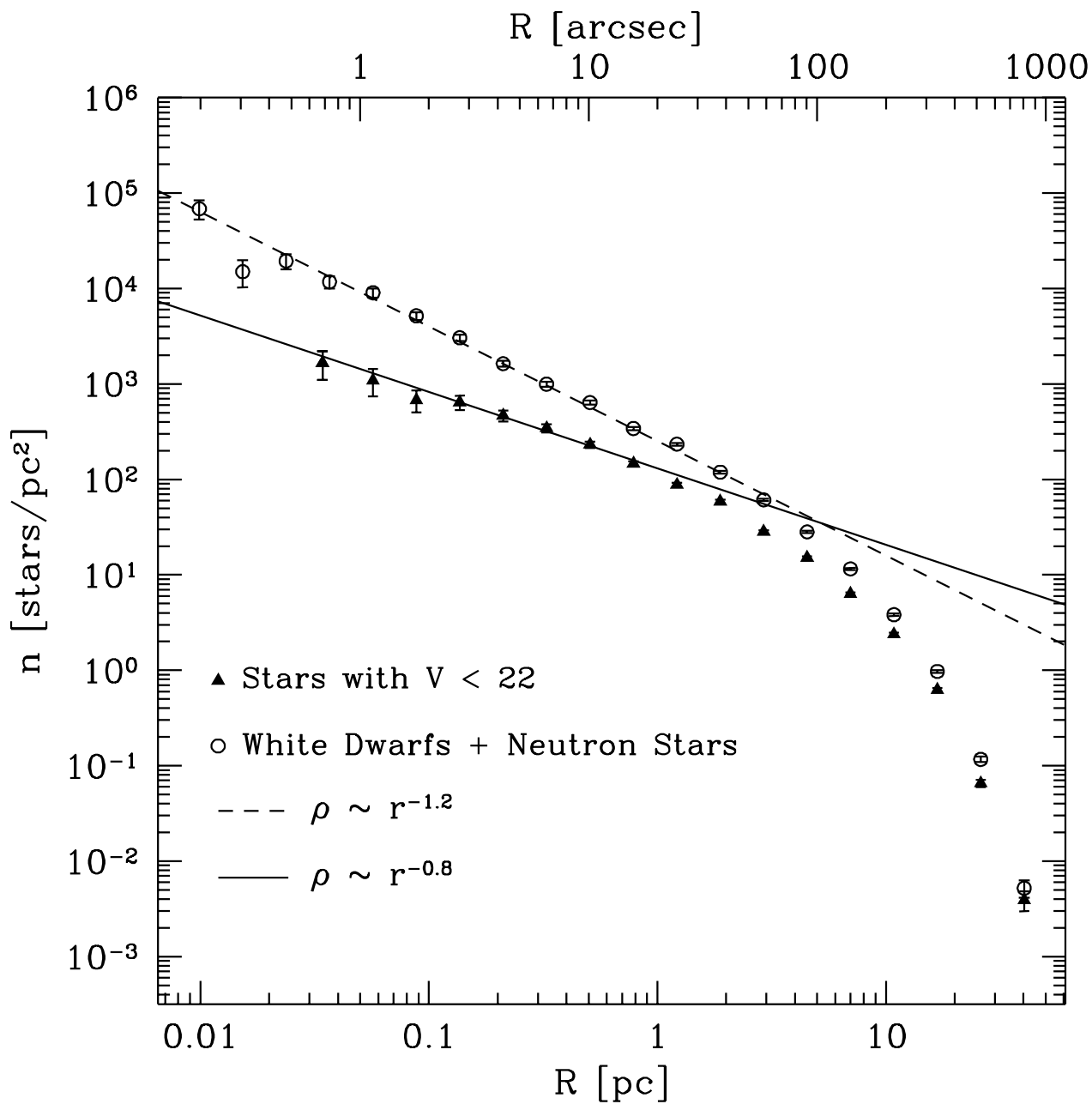}
\caption{Radial surface number density profiles for different stellar
groups just after core collapse.  Open circles denote white dwarfs and
neutron stars.  Filled triangles denote stars with $V<22$.  In the
center, the slope for bright stars is similar to that observed in M15.
The bright stars follow a much shallower distribution than the compact
remnants, due to mass segregation.
\label{fig2}}
\end{figure}

Figure \ref{fig2} depicts the surface number density of bright
($V<22$) stars and of compact remnants.  The adopted cutoff of $V=22$
is the photometric limit found in the study of the cluster center by
\citet{SosinKing1997}, and is also consistent with the limit of
$V=22.5$ in the data of \citet{vandermareletal2002}.  For both groups,
the inner region shows clear power-law cusps, with indices of
approximately $-0.8$ and $-1.2$, respectively.  The surface density of
bright stars is again in very good agreement with the HST WFPC2 and
FOC star count results \citep{Guhathakurtaetal1996,SosinKing1997}.

A cluster in deep collapse should have a density profile steeper than
isothermal, since the velocity dispersion increases inward
\citep{LBE1980, Cohn1980}.  One might wonder why the central density
profile of bright stars shows a slope shallower than that of an
isothermal sphere ($\sim r^{-1}$ in projection).  The reason is simply
that the bright stars are not the most massive components in
present-day globular clusters \citep{MurphyCohn1988, Luggeretal1995}.
Compact remnants (neutron stars and massive white dwarfs) are more
massive, and are the dominant population in the central region (see
Figure\,\ref{fig2}).  Their density profile ($\rho \sim r^{-2.2}$ in 3
dimensions) is close to the theoretical prediction for the central
profile of a core-collapsed cluster: $\rho \sim r^{-2.26}$ (see
\citet{Baumgardtetal2002} and references therein).  As a consequence,
proper interpretation of Figure\,\ref{fig1} must take into account the
substantial radial variation of the mass to light ratio in the cluster
core.

\begin{figure}[htbp!]
\plotone{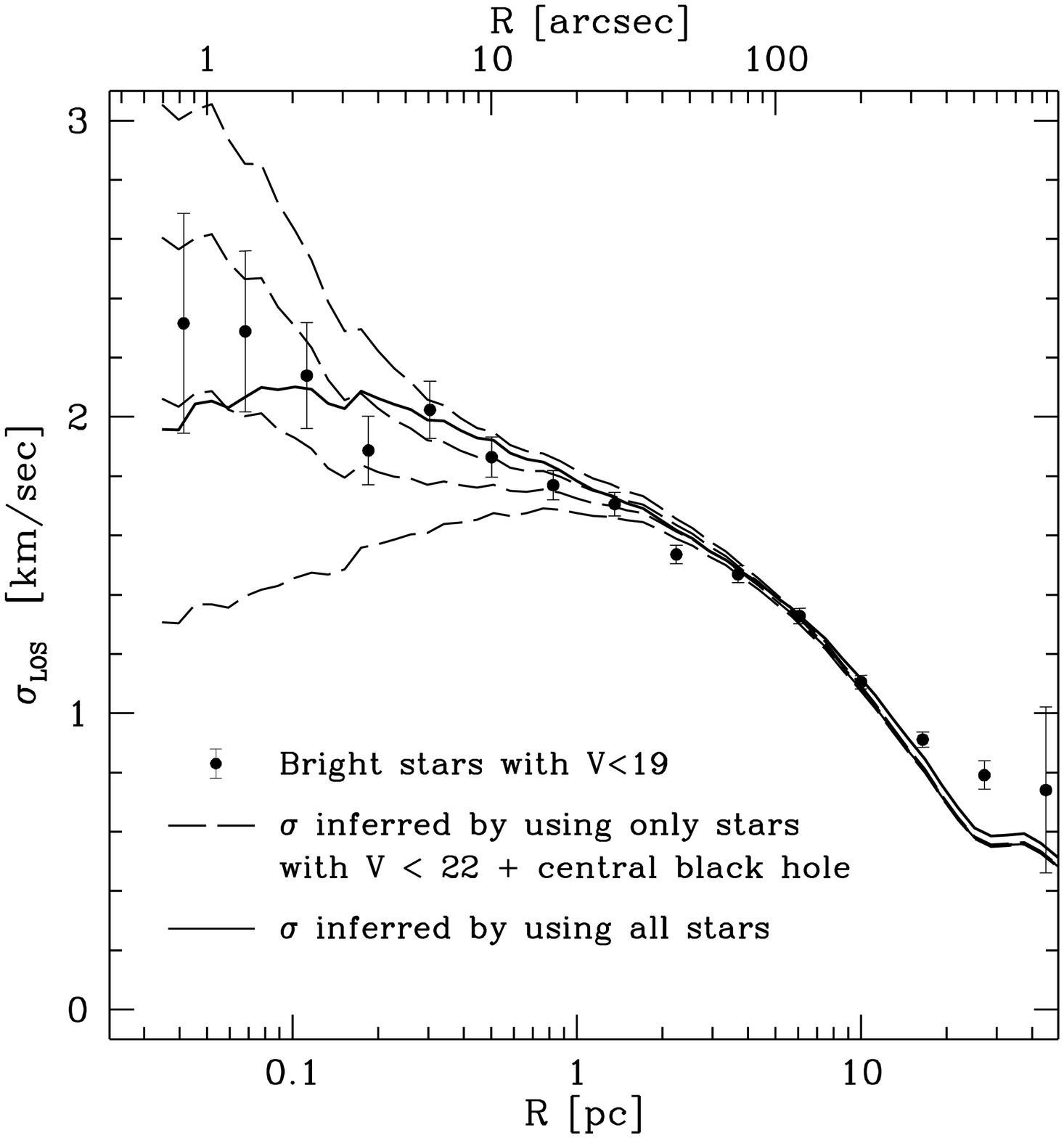}
\caption{Line-of-sight velocity dispersion of the $V<19$ stars in the
$N$-body simulations (filled circles), and inferred from the stellar
number density and cluster potential (solid and dashed curves).  The
solid curve shows the inferred velocity dispersion of stars with
$V<22$, using the potential calculated from all stars.  Dashed curves
are calculated using the potential determined from stars with $V<22$,
assuming a constant $M/L$, together with central point masses of
(bottom to top) 0, 40, 80 and 120 $M_\sun$.  The value of $M/L$ is
chosen to fit the measured velocity dispersion between 1 and 10 pc
from the cluster center.  For constant assumed $M/L$, the best fit has
$M_{\rm BH} \sim 80M_\sun$.
\label{fig3}}
\end{figure}

As an illustration, we compare the observed velocity dispersion
profile of our model with the velocity dispersion profile inferred
from the distribution of bright stars, using the Jeans equation with
an isotropic velocity distribution and a constant mass-to-light ratio.
The numerical procedure is as described by \citet{Gerssenetal2002}
(section 5).  Figure \ref{fig3} shows the result.  Not surprisingly,
we find a large discrepancy between the inferred velocity dispersion
and the observed profile, as illustrated by the lowest dashed line in
Figure 3.  The predicted central velocity dispersion, based on the
mass contribution of the visible stars, would actually dip in the
center, contrary to what is observed.  Most of the discrepancy is
caused by the neglect of the central concentration of dark matter in
the form of stellar remnants.  Trying to improve the fit by
introducing a central point mass as a free parameter leads to a
central mass of approximately 80 $\mbox{M}_{\odot}$ (second dashed
line from the top in Figure\,\ref{fig3}).

\citet{Gerssenetal2002} have analysed the velocity distribution of the
bright stars in M15, using two different methods and averaging the
results.  They first assume a constant mass-to-light ratio, then adopt
a more realistic radial run of mass to light, obtained from
Fokker-Planck simulations \citep{Dulletal1997}.  Their first method
leads to an inferred central black hole mass of $3.2 \times 10^3 \;
\mbox{M}_\odot$, containing a fraction of $3.2 \times 10^3 \;
\mbox{M}_\odot/4.9 \times 10^5 \; \mbox{M}_\odot = 0.65\%$ of the
total cluster mass.  (The choice of M15 mass is taken from Dull {\em
et al.})  This is similar to the fractional mass of the central point
mass deduced above from Figure \ref{fig3}, to which we ascribed a mass
ratio of $80 \; \mbox{M}_\odot/20,000 \; \mbox{M}_\odot = 0.4\%$ of
our cluster mass.

Using the correct cluster potential (solid line in Figure\,\ref{fig3})
in the analysis of our simulations recovers the velocity dispersion of
stars of $V<19$ without the need for a central point mass.  In effect,
we use the (known) variation in the mass-to-light ratio of the model
cluster to convert from the observed $V<22$ number density to the
actual potential.  Comparison of the central point-mass data with the
error bars in the ``observed'' ($V<19$) velocity dispersion in
Figure\,\ref{fig3} suggests that the largest point mass that could be
hidden in the data has a mass of $\aplt 40 M_\odot$.  This would
correspond to $\aplt 10^3 M_\odot$ in the M15 system.  
In contrast, the equivalent (second) method employed by
\citet{Gerssenetal2002} actually increased the inferred central mass
to $4.5 \times 10^3 \; \mbox{M}_\odot$.  This mass determination was
subsequently shown to be erroneous \citep{Gerssenetal2003}.
\footnote{\citet{Gerssenetal2003} report that Figures 9 and 12 of Dull
{\em al.} contained errors which critically affected their analysis.
They were already aware of this fact, and informed us of it after
receiving a copy of the submitted manuscript of the current paper.}
The correct treatment yielded a formal mass of $1.7 \times 10^3 \;
\mbox{M}_\odot$; however a mass of zero was excluded only at the
$\sim\frac12\,\sigma$ level.

In their addendum, \citep{Gerssenetal2003} maintain that a central
black hole remains a viable interpretation of the M15 data, citing the
probability that most neutron stars would have escaped the cluster on
formation, in contradiction to the assumption of 100\% neutron star
retention made by \citet{Dulletal1997} and also in Figure 3 above.
Most neutron stars receive substantial ``kicks'' at birth
\citep{LyneLorimer1994}, which may eject them from their parent
cluster.  Theoretical estimates of the retention fraction range from
$\sim5$ to $\sim$20\% \citep{Drukier1996}.  If no neutron stars were
present in the core, the slope of the luminosity profile would be
expected to steepen somewhat \citep{TakahashiLee2000}.

To address this possibility, we have repeated our earlier simulation
with the extreme alternative assumption that no neutron stars were
retained.  The result is plotted in Figure 4, which presents the
analogous information to Figure 3 for this model.  The discrepancy
between the ``observed'' velocity profile and the expected profile
calculated from the distribution of stars with $V<22$, assuming a
constant mass-to-light ratio, still exists, since in this case massive
white dwarfs have accumulated in the center and replaced the
main-sequence stars.  Thus, changing the neutron star retention
fraction does not significantly alter our conclusion.  The assumption
of constant $M/L$ now yields a black-hole mass of $\sim40M_\sun$, half
the value found in Figure 3, since massive white dwarfs and neutron
stars contributed roughly equally to the central dark mass in that
model.

Using the same reasoning as before, we then estimate a value of
$20M_\sun$, for the maximum mass that could be hidden in the form of a
central black hole.  In the case of M15, this would correspond to
$\aplt 500 M_\odot$.  We note, however, that the fitting procedure is
relatively insensitive to the precise nature of the dark matter
contained within the innermost 0.5 pc (H. Cohn and P. Lugger, 2002,
private communication).  The present data are probably consistent with
dark matter in the form of a range of combinations of neutron stars,
massive white dwarfs, or an intermediate-mass black hole.

\begin{figure}[htbp!]
\plotone{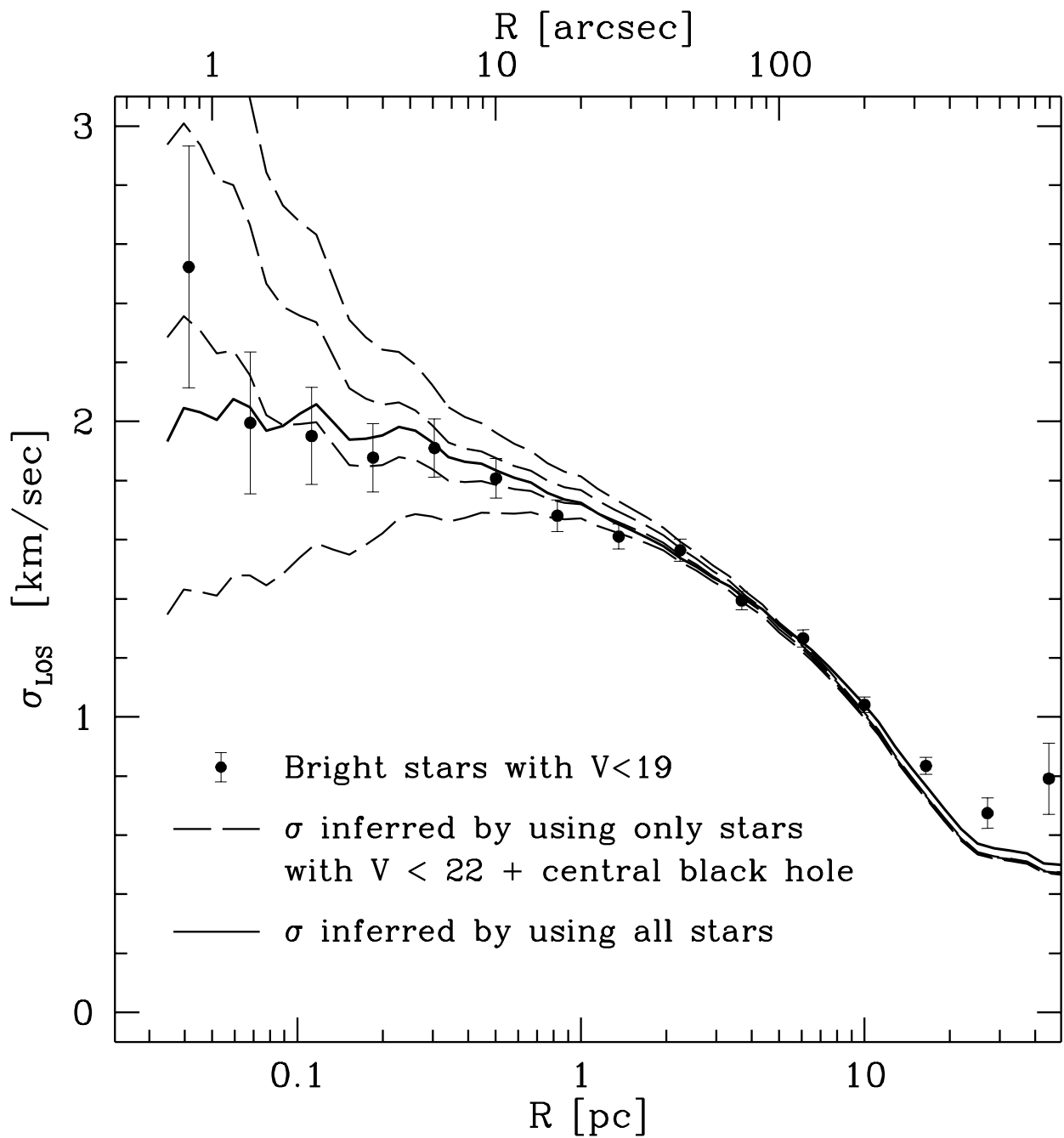}
\caption{As for Figure 3, but for a model with 0\% neutron star
retention.  For constant $M/L$, the best-fitting black-hole mass is
now $40M_\sun$.
\label{fig4}}
\end{figure}

\section{Conclusions}

In this paper we compare recent observations of the central regions of
M15 with recent direct $N$-body simulations of realistic models of
star clusters.  We find that the velocity dispersion and luminosity
profiles obtained from the $N$-body simulations, after appropriate
scaling, reproduce the observations without any need to invoke a
central point mass.  Earlier Fokker--Planck results without a black
hole \citep{Dulletal1997, Dulletal2002} are also consistent with the
current observations.  Thus we conclude that the M15 observations can
be adequately explained without recourse to a central massive black
hole.

Although the current observations do not prove the existence of a
central massive black hole, they do not disprove it either.  Our
analysis (Figure\,\ref{fig3}) indicates that a moderate
intermediate-mass black hole of $\sim10^3$ solar masses is still
possible.  Such an object is not altogether unexpected, since it might
have formed early in the cluster's evolution through runaway merging
\citep{Ebisuzakietal2001,PortegiesZwartMcMillan2002}.  To confirm this
interesting possibility, or to place more stringent limits on the mass
of a possible black hole, will require detailed evolutionary modeling
of the cluster for different evolutionary scenarios.  We plan to carry
out such simulations in the near future.

\section*{Acknowledgments}

HB and JM thank Toshi Fukushige and Yoko Funato for stimulating
discussions; PH and SM acknowledge several collegial conversations
with Roeland van der Marel, Karl Gebhardt, Haldan Cohn and Phyllis
Lugger following submission of this paper.  We also acknowledge
detailed and helpful comments on the manuscript by Phyllis Lugger and
an anonymous second referee.  This work is supported in part by
Grant-in-Aid for Scientific Research B (13440058) of the Ministry of
Education, Culture, Sports, Science and Technology, Japan, by grants
NASA ATP grants NAG5-6964 and NAG5-9264, and by the Royal Netherlands
Academy of Sciences (KNAW) and the Netherlands Research School for
Astronomy (NOVA)


\begin{thebibliography}{}

\bibitem[Aarseth(1999)]{Aarseth1999a}
Aarseth, S. J. 1999, \pasp, 111, 1333
\bibitem[Baumgardt et al.(2002)]{Baumgardtetal2002}
Baumgardt, H., Heggie, D., Hut, P., and Makino, J. 2002, submitted to \mnras
\bibitem[Baumgardt and Makino(2002)]{BaumgardtMakino2002}
Baumgardt, H., and Makino, J. 2002, to appear in {\mnras} ({\tt
astro-ph/xxxxx})
\bibitem[Bettwieser and Sugimoto(1984)]{Bettwieser1984}
Bettwieser, E., and Sugimoto, D. 1984, \mnras, 208, 493
\bibitem[Cohn(1980)]{Cohn1980}
Cohn, H. 1980, \apj, 242, 765
\bibitem[Djorgovski and Meylan(1994)]{DjorgovskiMeylan1994}
Djorgovski, S., and Meylan, G. 1994, \aj, 108, 1292
\bibitem[Drukier(1996)]{Drukier1996}
Drukier, G. A. 1996, \mnras, 280, 498
\bibitem[Dull et al.(1997)]{Dulletal1997}
Dull, J. D., Cohn, H. N., Lugger, P. M., Murphy, B. W., Seitzer, P. O.,
 Callanan, P. J., Rutten, R. G. M., and Charles, P. A. 1997, \apj, 481, 267
\bibitem[Dull et al.(2002)]{Dulletal2002}
Dull, J. D., Cohn, H. N., Lugger, P. M., Murphy, B. W., Seitzer, P. O.,
 Callanan, P. J., Rutten, R. G. M., and Charles, P. A. 2002, submitted
to {\apj} ({\tt astro-ph/0210588})
\bibitem[Ebisuzaki et al.(2001)]{Ebisuzakietal2001}
Ebisuzaki, T., Makino, J., Tsuru, T. G., Funato, Y., Portegies Zwart, S. F.,
 Hut, P., McMillan, S. L. W., Matsushita, S., Matsumoto, H., and Kawabe, R.
 2001, \apjl, 562, L19
\bibitem[Gerssen et al.(2002)]{Gerssenetal2002}
Gerssen, J., van der Marel, R. P., Gebhardt, K., Guhathakurta, P.,
Peterson, R., and Pryor, C. 2002, \aj, in press
\bibitem[Gerssen et al.(2003)]{Gerssenetal2003}
Gerssen, J., van der Marel, R. P., Gebhardt, K., Guhathakurta, P.,
Peterson, R., and Pryor, C. 2003, \aj, in press
\bibitem[Goodman and Hut(1989)]{GoodmanHut1989}
Goodman, J., and Hut, P. 1989, \nat, 339, 40
\bibitem[Guhathakurta et al.(1996)]{Guhathakurtaetal1996}
Guhathakurta, P., Yanny, B., Schneider, D. P., and Bahcall, J. N. 1996,
 \aj, 111, 267
\bibitem[Harris(1996)]{Harris1996}
Harris, W. E. 1996, \aj, 112, 1487 ({\tt
http://physun.physics.mcmaster.ca/Globular.html})
\bibitem[Hurley et al.(2000)]{Hurleyetal2000}
Hurley, J. R., Pols, O. R., and Tout, C. A. 2000, \mnras, 315, 543
\bibitem[Kroupa(2001)]{Kroupa2001}
Kroupa, P. 2001, \mnras, 322, 231
\bibitem[Lee(1987)]{Lee1987}
Lee, H. M. 1987, \apj, 319, 801
\bibitem[Lugger et al.(1995) ]{Luggeretal1995}
Lugger, P. M., Cohn, H. N., \& Grindlay, J. E. 1995, \apj, 439, 191
\bibitem[Lynden-Bell \& Eggleton(1980)]{LBE1980}
Lynden-Bell, D., \& Eggleton, P. P. 1980, \mnras, 191, 483
\bibitem[Lyne \& Lorimer(1994)]{LyneLorimer1994}
Lyne, A. G., Lorimer, D. R. 1994, Nature, 369, 127
\bibitem[Makino(1997)]{Makino1997}
Makino, J. 1997, \apj, 478, 58
\bibitem[Makino et al.(2002)]{Makinoetal2002}
Makino, J., Fukushige, T., and Namura, K. 2002, in preparation
\bibitem[van der Marel et al.(2002)]{vandermareletal2002}
van der Marel, R. P., Gerssen, J., Guhathakurta, P., Peterson, R. C., and
Gebhardt, K. 2002, \apj, in press
\bibitem[Meylan and Heggie(1997)]{MeylanHeggie1997}
Meylan, G., and Heggie, D. C. 1997, \aapr, 8, 1
\bibitem[Miller and Hamilton(2002)]{MillerHamilton2002}
Miller, M. C., and Hamilton, D. P. 2002, \mnras, 330, 232
\bibitem[Murphy \& Cohn(1988)]{MurphyCohn1988}
Murphy, B. W., \& Cohn, H. N. 1988, \mnras, 232, 835
\bibitem[Portegies Zwart et al.(1999)]{PortegiesZwartetal1999}
Portegies Zwart, S. F., Makino, J., McMillan, S. L. W., and Hut,
P. 1999, \aap, 348, 117
\bibitem[Portegies Zwart et al.(2001)]{PortegiesZwartetal2001}
Portegies Zwart, S. F., McMillan, S. L. W., Hut, P., and Makino,
J. 2001, \apj, 546, L101
\bibitem[Portegies Zwart and McMillan(2002)]{PortegiesZwartMcMillan2002}
Portegies Zwart, S. F., and McMillan, S. L. W. 2002, \apj, 576, 899
\bibitem[Sosin and King(1997)]{SosinKing1997}
Sosin, C., and King, I. R. 1997, \aj, 113, 1328
\bibitem[Spitzer(1987)]{Spitzer1987}
Spitzer, L. J. 1987, {\em Dynamical Evolution of Globular Clusters}.
 Princeton University Press, Princeton, New Jersey.
\bibitem[Takahashi \& Lee(2000)]{TakahashiLee2000}
Takahashi, K. and Lee, H. M.  2000, \mnras, 316, 671

\end{thebibliography}
\end{document}